\documentclass[aps,prl,twocolumn,superscriptaddress]{revtex4} 
\usepackage[T1]{fontenc}
\usepackage{amssymb}

\usepackage{hyperref}
\hypersetup{
    bookmarks=true,         
    unicode=true,          
    pdftoolbar=false,        
    pdfmenubar=true,        
    pdffitwindow=true,      
    pdftitle={My title},    
    pdfauthor={Author},     
    pdfsubject={Subject},   
    pdfnewwindow=true,      
    pdfkeywords={keywords}, 
    colorlinks=true,       
    linkcolor=red,          
    citecolor=black,        
    filecolor=magenta,      
    urlcolor=cyan           
}

\usepackage[all]{hypcap}
\usepackage{graphicx}
\usepackage{amsmath}
\usepackage{natbib}
\usepackage{layouts}

\begin{document}

\title{Exchange Interactions in Paramagnetic Amorphous and Disordered Crystalline CrN-based Systems}

\author{A. Lindmaa} 
\email{alexander.lindmaa@liu.se}
\affiliation{Department of Physics, Chemistry and Biology (IFM), Link\"oping University, SE-581 83  Link\"oping, Sweden}
\author{R. Liz\'arraga}
\affiliation{Instituto de Ciencias F\'isicas y Matem\'aticas, Facultad de Ciencias, Universidad Austral de Chile, Casilla 567, Valdivia, Chile}
\author{E. Holmstr\"om}
\affiliation{Instituto de Ciencias F\'isicas y Matem\'aticas, Facultad de Ciencias, Universidad Austral de Chile, Casilla 567, Valdivia, Chile}
\author{I. A. Abrikosov}
\affiliation{Department of Physics, Chemistry and Biology (IFM), Link\"oping University, SE-581 83 Link\"oping, Sweden}
\author{B. Alling}
\affiliation{Department of Physics, Chemistry and Biology (IFM), Link\"oping University, SE-581 83 Link\"oping, Sweden}

\date{\today}

\begin{abstract} 
We present a first principles supercell methodology for the calculation of
exchange interactions of magnetic materials with arbitrary degrees of structural
and chemical disorder in their high temperature paramagnetic state. It is based
on a projection of the total magnetic energy of the system onto local pair
clusters, allowing the interactions to vary independently as a response to their
local environments. We demonstrate our method by 
deriving the distance dependent
exchange interactions in vibrating crystalline CrN, a  Ti$_{0.5}$Cr$_{0.5}$N solid
solution as well as in amorphous CrN.  
Our method reveals strong local environment effects in all three systems. In the amorphous case we use
the full set of exchange interactions in a search for the non-collinear magnetic ground
state.

\end{abstract}

\maketitle 

The understanding and accurate modeling of magnetic materials at finite
temperatures is a grand challenge in solid state physics~\cite{Moriya1985}.
Its importance is highlighted by demands for new iron-based materials in the
steel industry, improvements in materials for electrical
motors and generators, and the outlook of
the full utilization of the spin degree of freedom in
electronics~\cite{Hasegawa2004, Sinova2012}. 
The difficulties in providing an accurate
description of high-temperature magnetic materials stem from the complexity of
the quantum excitations that must be included in the models and simulations. Such
excitations are of electronic, magnetic, vibrational, and structural nature with unknown individual impact.

Particularly, the treatment of amorphous magnets is problematic, as the lack
of crystal symmetry hampers both experimental characterization as well as
drastically increases computational costs in the theoretical approaches~\cite{Kazimirov2008,Hostert2012}. Furthermore, the topological disorder is likely to induce unintuitive
non-collinear magnetic configurations and methodological development is needed to understand their excitations.

In
crystalline cases, one key approach to get around the quantum complexity, 
has been the mapping of the magnetic configurational degree
of freedom onto a semi-classical model Hamiltonian such as the Heisenberg
model known for its applicability on systems with robust local moments. This model Hamiltonian,

\begin{equation} \label{eq:heiss}
\mathcal{H}=-\sum_{i \neq j}J_{ij} \boldsymbol{e}_i,\boldsymbol{e}_j ,
\end{equation} 

\noindent
where the $J_{ij}$'s are the magnetic exchange interactions (MEI)
between pairs of magnetic atoms $(i,j)$ and each $\boldsymbol{e}_i$ is a unit
vector directed along the local atomic moment at site $i$,  can subsequently be used in Monte
Carlo~\cite{Ekholm2010} or spin dynamics
simulations~\cite{Hellsvik2008,Skubic2008} to obtain critical temperatures for ordering (T$_C$), or to find ground state configurations.

This Heisenberg Hamiltonian, with only bi-linear pair interactions gives only an approximate description of the real complex magnetism in solid state systems. Its interactions correspond to the first terms of the complete expansion series of the magnetic configurational energy around the value of the fully disordered magnet. It neglects interactions corresponding to multi-site clusters, bi-quadratic terms and other higher order terms in the expansion. Nevertheless it has proven to be valuable in practice and is known to give an accurate description of the magnetic energies of several crystalline systems. Certainly, limitations of this magnetic Hamiltonian should be kept in mind.


Several mapping procedures based on first principles electronic structure
calculations have been successfully employed to obtain the MEI. The
perturbative magnetic force theorem proposed by Liechtenstein \emph{et al.}
\cite{Liechtenstein,Liechtenstein1984} has played an important role and has been
implemented together with the disordered local moments (DLM)~\cite{Gyorffy1985}
treatment of paramagnetism in the generalized perturbation method (GPM)
\cite{Ruban2004}. Supercell approaches include the frozen magnon approach~\cite{rosengaard97,PhysRevB.76.184406} and the structure inversion methodology~\cite{Singer2011}. 
Despite their success, all those methods suffer from difficulties to treat
systems with arbitrary disorder as they, this far, have relied on the existence of
an underlying lattice geometry. Hence, in our opinion, there is no established approach to obtain paramagnetic MEIs in systems with
topological disorder, or even in configurationally disordered
crystalline alloys with large local lattice relaxations. 
Moreover, the impact of temperature induced vibrations, the need to derive MEI  from a disordered magnetic reference state~\cite{Ruban2004,Alling2009}, and local environment effects are often neglected.

In this work we propose a magnetic direct cluster averaging (MDCA) method, to
calculate paramagnetic exchange interactions in a system with arbitrary
geometry. The method is based on a
conventional first principle supercell approach treating crystalline and
non-crystalline materials on an equal footing. We illustrate the
method for crystalline disordered rock-salt structure CrN and Ti$_{0.5}$Cr$_{0.5}$N as well as
amorphous CrN. 
The studied materials have attained substantial attention due to couplings
between magnetic, electronic, and structural parameters with significant implications for
technological applications~\cite{Filippetti2000,Rivadulla2009,
Alling2010natmat,Bhobe2010,Wang2012}.
We also demonstrate that
a benefit from the MDCA method is that it can be used as a baseline to ensure convergence
within the Connolly-Williams (CW) \cite{Connolly1983} structure inversion method.

The Hamiltonian in Eq.~\ref{eq:heiss} describes both collinear and non-collinear
magnetic configurations. Under the assumption that there are no important non-linear higher order terms present, such as the ones discussed in Ref~\cite{Drautz2005}, the $J_{ij}$ can be obtained from calculations restricted to
collinear configurations.
This observation is especially valid when the
interactions in the high temperature paramagnetic state are desired, where the spin
correlation functions are small and one can follow the philosophy of the DLM
approach~\cite{Gyorffy1985,PhysRevB.82.184430}. 
Thus, to obtain the exchange interactions we consider a system with a collinear magnetic configuration specified as $\boldsymbol{\sigma}= \{ \sigma_1, \sigma_2, \ldots, \sigma_n \},$ 
where $\sigma_i \in \{ \pm 1\}$ are spin variables for each site with a
magnetic atom. This picture is equivalent to that of the configurational aspects
of a binary alloy for which a general mathematical framework
were developed by Sanchez \emph{et al.} \cite{1984PhyA..128..334S}. 
Following
this analogy the magnetic interactions may be viewed as effective pair interactions in a cluster
expansion procedure.
However, since our aim is to calculate magnetic
interactions in \emph{disordered} systems, \emph{where all pairs of atoms are unique}, we
can not rely directly on the traditional structure inversion approach, as it requires the presence of an underlying lattice
symmetry to reduce the number of free fitting parameters. Instead we start with
the very definition of the pair interactions as projections of the total energy, $E$,
onto the cluster basis function of each individual pair of atoms. Accordingly,
these projections define the exchange interactions, $J^{*}_{ij}$, as

\begin{equation}\label{eq:J*}
J^{*}_{ij}=\frac{1}{N_{\boldsymbol{\sigma}^\prime}}\sum_{\boldsymbol{\sigma}^\prime}\bigg[-\frac{1}{8}\!\sum_{\sigma_i,\sigma_j=\pm1}\! \! \!E\big(\{ \sigma_i, \sigma_j \} ; \boldsymbol{\sigma}^\prime \big)\prod_{k=i,j}\sigma_k\bigg],
\end{equation}

\noindent
where the summation is performed over all possible configurations
$\boldsymbol{\sigma}$ of $N$ magnetic atoms divided in a summation over the configuration within the
cluster $\{ \sigma_i, \sigma_j \}$ and in the rest of the cluster denoted
$\boldsymbol{\sigma}^\prime$. $N_{\boldsymbol{\sigma}^\prime} = 2^{(N-2)}$ is the total number of
collinear magnetic configurations of the remaining $(N-2)$ sites. 
%
%
This procedure does have a
counterpart in alloy theory  that was suggested by Berera~\cite{Berera1988} and
is referred to as direct cluster averaging (DCA).
Here, we define the expression within the brackets of
Eq. \ref{eq:J*} as two-site magnetic potentials, according to

\begin{eqnarray}\label{eq:Jij_sigmaprime}
J^{\boldsymbol{\sigma}^\prime}_{ij}&=& -\frac{1}{8}\sum_{\sigma_i, \sigma_j=\pm1}E\big(\{ \sigma_i, \sigma_j \} ; \boldsymbol{\sigma}^\prime \big)\prod_{k=i,j}\sigma_k  \\ \nonumber
&=& -\frac{1}{8} \bigg[   E\big(\{ \uparrow, \uparrow \} ; \boldsymbol{\sigma}^\prime \big)  + E\big(\{ \downarrow, \downarrow\} ; \boldsymbol{\sigma}^\prime \big) \\ \nonumber 
&-&E\big(\{ \uparrow, \downarrow\} ; \boldsymbol{\sigma}^\prime \big) -E\big(\{
\downarrow, \uparrow\} ; \boldsymbol{\sigma}^\prime \big) \bigg].
\end{eqnarray}

\begin{figure}[htb] 
	\centering
 	\includegraphics[height=1.00\columnwidth, angle=270]{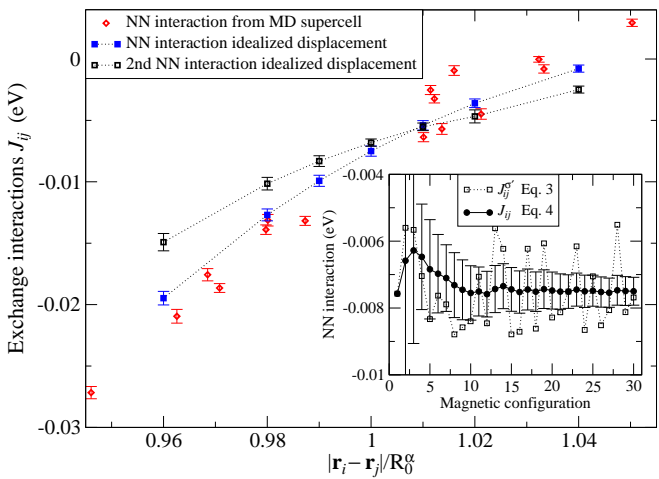}
  	\caption{(Color online) The calculated magnetic exchange
interactions on the first two Cr-Cr coordination shells of cubic CrN as a function of
the interatomic distance. The values obtained from geometries derived with
DLM-MD as well as idealized displacements of the atoms are shown. $\mathrm{R_0^{\alpha}}$ is the equilibrium pair distance of the $\alpha$:th coordination shell ($\alpha=1,2)$. The inset illustrates the application of Equations~\ref{eq:Jij_sigmaprime} and~\ref{eq:Jij}.}
 	\label{fig:Jij_r}
\end{figure}

\noindent
We note that by including all the four energy terms in Eq.~\ref{eq:Jij_sigmaprime} the pair interaction between the moments on sites $i$ and $j$ is singled out as the effect of any other pair interaction between one of the moments within the cluster and any moment outside it is cancelled. The averaging over the configurations $\boldsymbol{\sigma}'$ in Eq~\ref{eq:J*} removes any possible effect of multisite interactions.
In practical calculations, restrictions have to be imposed on the number of configurations that
are considered, introducing an uncertainty in the values of $J_{ij}^{*}$.
However if we choose a subset of $N_{\boldsymbol{\zeta}'}$ configurations $\{{\boldsymbol{\zeta}'}\}
\subset \{\boldsymbol{\sigma}^\prime\}$ randomly, we simultaneously obtain paramagnetic-like configurations and allow for a treatment of the $J^{\boldsymbol{\sigma}^\prime}_{ij}$ as stochastic variables. 
Thus we can estimate the MEI

\begin{equation}\label{eq:Jij}
J_{ij}^{*}\approx J_{ij}=\frac{1}{N_{\boldsymbol{\zeta}'}}\sum_{\boldsymbol{\sigma}'\in \{{\boldsymbol{\zeta}'}\}} J^{\boldsymbol{\sigma}'}_{ij},
\end{equation}

\noindent
with a confidence interval of degree $1-\alpha$ with respect to the statistical sampling of magnetic configurations outside the pair: 
$J_{ij}-t_{\alpha/2}(f)d<J^{*}_{ij}<J_{ij}+t_{\alpha/2}(f)d$, 
where $t_{\alpha/2}(f)$ is a $t-$distribution with $f=N_{\boldsymbol{\zeta}'}-1$ degrees of freedom and $d=s/\sqrt{N_{\boldsymbol{\zeta}'}}$ where $s$ is the estimated standard deviation of the $J_{ij}$. In this work we have used 95\% confidence intervals plotted as error bars in the figures. The inset of Fig.~\ref{fig:Jij_r} shows the individual two-site potentials (Eq.~\ref{eq:Jij_sigmaprime}) and the accumulated estimate  of $J_{ij}^{*}$ (Eq.~\ref{eq:Jij}) for the case of nearest neighbor MEI  in ideal CrN. In this case the set of configurations $\{{\boldsymbol{\zeta}'}\}$ were based on 30 randomly generated supercells with an average spin correlation function on the first coordination shell as small as -0.006. 



The electronic structure problem was solved using density functional theory
(DFT)~\cite{Hohenberg1964,Kohn1965} and the projector augmented wave (PAW) method
\cite{PhysRevB.50.17953}, as implemented in the Vienna \emph{ab-initio}
simulation package (VASP) \cite{PhysRevB.48.13115, PhysRevB.59.1758}. To
accurately describe strongly correlated Cr 3d-electrons, the local spin density
approximation with an additional Hubbard U-term (LDA+U) \cite{PhysRevB.44.943,
PhysRevB.57.1505} of $3~eV$ was employed as discussed in details in
Ref.~\cite{PhysRevB.82.184430}. 64 atom supercells were used for the
derivation of NN interactions in CrN while 96 atoms cells were used to obtain
next-NN interactions in CrN and the interactions in Ti$_{0.5}$Cr$_{0.5}$N. 250
atoms were used in the simulations of the amorphous CrN structure.  The size of our calculation
cells were carefully checked so that effects from interactions between periodic
images were negligible. 

The first application of the method is to investigate the influence of lattice
vibrations on the magnetic exchange interactions. Fig.~\ref{fig:Jij_r} shows the
calculated (Eq.~\ref{eq:Jij_sigmaprime} and~\ref{eq:Jij}) nearest neighbor (NN)
interactions in a realistic finite temperature geometry obtained in
Ref.~\cite{Steneteg2012} by the disordered local moment molecular dynamics
(DLM-MD) at 300 K.  The figure shows interactions derived between the same pair
of atoms at 16 different MD-time steps, thus with different inter-atomic
distances. To visualize the pure distance effect in $J_{ij}(|\mathbf{r}_{i}-\mathbf{r}_j|)$ we also
calculated the values of the nearest and next-nearest neighbor interactions,
when only the interaction pair were distorted from ideal lattice positions. 


We observe a distinct dependency of the MEI on the
distances between the atoms, even for slight deviations. The NN as well as the next-NN
interactions are antiferromagnetic at zero-deviation (-7.5 meV and -6.8 meV
respectively) with a sharp increase in antiferromagnetic (AFM) strength at smaller distances. However, when positioned far
away, but still at realistic distance as illustrated with presence in the
MD-data,  the NN interaction actually does become positive. This qualitative
difference between magnetic interactions induced by $300~K$ vibrations opens
perspectives of phonon induced dynamic magnetic short range correlations in the
paramagnetic state, and possibly, the opposite coupling depending on the magnetic and vibrational timescales.

\begin{figure}[htb]
	\centering
 	\includegraphics[height=1.00\columnwidth, angle=270]{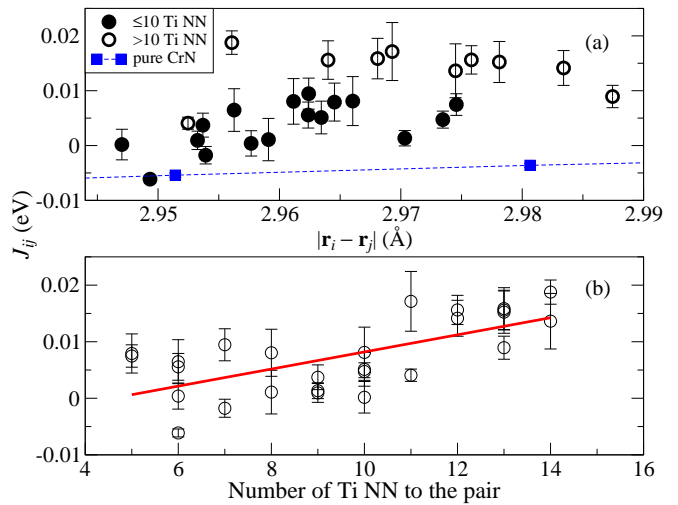}
	\caption{(Color online) The magnetic exchange interactions between Cr-Cr
                  pairs in a Ti$_{0.5}$Cr$_{0.5}$N alloy as a function of the interatomic
                  distances (top panel) and number of Ti nearest neighbors to the Cr-Cr
                  pairs (lower panel). In the top panel the Ti rich environments are highlighted by
                  open symbols and the values in pure CrN are included for comparison. In the lower
                  panel, a linear regression trend line is provided as a guide for the eye.} 
 	\label{fig:Jij}
\end{figure}

\begin{figure}[htb]
	\centering
 	\includegraphics[height=1.0\columnwidth, angle=270]{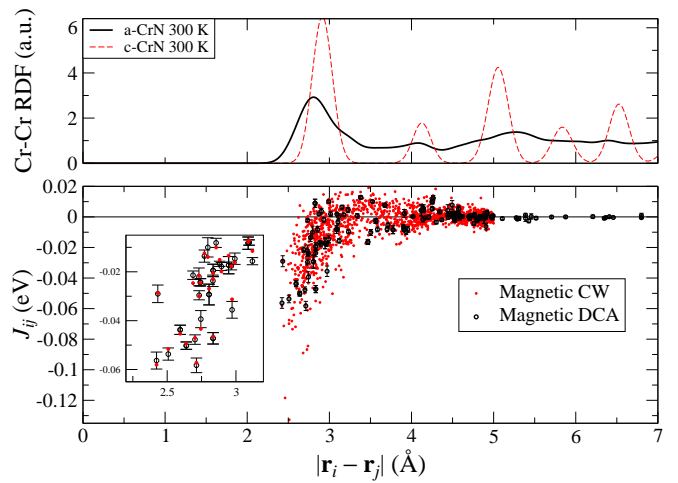}
	\caption{ (Color online) The Cr-Cr radial distribution function for the
                   amorphous CrN model (a-CrN) compared to crystalline cubic CrN (c-CrN) (top
                   panel) and the magnetic exchange interactions (lower panel) in the amorphous
                   cell obtained with MDCA and MCW approaches respectively. The inset in the lower
                   panel is a zoom in of the short distance interactions comparing the MDCA results
                   with the same interactions obtained with MCW.}
 	\label{fig:aCrN} 
\end{figure}

The second application of the method is to study effects from substitutional
chemical disorder in solid solutions. In such systems, different local chemical
environments give rise to lattice relaxations and differences in the electronic
structure of the magnetic components, and thus possibly differences in the
exchange interactions. Such effects are for instance crucial to understand the 
INVAR-effect in FeNi alloys~\cite{Ruban2005,Liot2009}.

Shown in Fig. ~\ref{fig:Jij}, are the exchange interactions between a
selection of Cr-Cr NN pairs on the metal fcc sublattice in a supercell model for the alloy
Ti$_{0.5}$Cr$_{0.5}$N~\cite{Alling2010TiCrN}, as a function of the
interatomic distance in \AA (top panel)
and as a
function of the number of Ti atoms that are NN to the pair (lower panel).  The exchange
interactions are generally ferromagnetic, in line with
experiments~\cite{Aivazov1975,Inumaru2007} but in sharp contrast to AFM CrN.
Cr-Cr pairs with more than 10 NN Ti atoms seem to exclusively interact
ferromagnetically. Thus, the increasing presence of Ti atoms in the immediate
surrounding of a Cr-Cr pair strongly influences their exchange coupling.  This
study reveals the importance of local environment effects on magnetic exchange
interactions in these alloys, adding further complexity to the concentration
dependence observed previously in this material~\cite{Alling2010TiCrN}. 

The third application of our method is to 
study an extreme case of disorder
in the form of topologically disordered amorphous CrN. We obtained our 
structural model for amorphous CrN by means of the stochastic quench
method~\cite{Holmstrom2009b, Holmstrom2010} using 250 atoms in the supercell. To
include the effect of temperature induced vibrations on the structure we first
run 10 000 time steps of 1 fs by means of a standard quantum molecular dynamics
method. The magnetic state was kept fixed and the temperature was 700 K,
a typical growth temperature for amorphous nitride thin
films~\cite{Zhou2011}. The QMD simulations are carried out using a canonical ensemble (NVT), neglecting thermal expansion. In order to maintain the temperature and avoid artificial energy drift we use the standard Nos\'e thermostat~\cite{Nose1991} implemented in VASP, with the Nos\'e-mass corresponding to a 40 time step period. Then we run 1400 time steps of 1 fs of DLM-MD~\cite{Steneteg2012} at 300 K
to average out any possible memory effects of specific magnetic orientations in the
geometry. 

The top panel of figure~\ref{fig:aCrN} shows the obtained Cr-Cr radial
distribution functions (RDF) for our amorphous model as well as for our
crystalline cubic CrN model after the MD simulation, both convoluted with a 0.2
\AA~gaussian. The first peak of the a-CrN Cr-Cr RDF is at slightly lower distances as compared to c-CrN as several of the Cr-atoms in the a-CrN case are positioned in a N-poor local environment allowing for smaller Cr-Cr distances. However, the volume of a-CrN is as expected larger than c-CrN.

In our supercell approach, we first investigated the range of 
the magnetic interactions. 
This was achieved by using the MDCA approach to calculate  a subset of 20 chosen interactions out
of the more than 4000 unique Cr-Cr pairs in our cell that has distances less
than half of the supercell periodicity. 
We found that Cr-Cr pairs with
larger separations than 5 \AA~have negligible interaction strengths of at most 1
meV, a value which is within our statistical error bars. This distance corresponds roughly to that of the 3rd metal coordination shell in a CrN crystal. Thus,
we could focus on the 1456 Cr-Cr pairs with interatomic
distances less than 5\AA, in our supercell of size
13.603 \AA. 

In the lower panel of Fig.~\ref{fig:aCrN} we show 127 exchange interactions
obtained with our MDCA method. 
Our approach gives swift access to any chosen individual pair interaction
but if all the thousands of interactions in the amorphous
supercell are desired, the method is computationally cumbersome. However,
the initial MDCA survey of the $J_{ij}$s has revealed both the relevant cut-off
and provided a substantial subset of reliable interaction values. This
translates into both a necessary limitation of the number of free parameters and
a reliable convergence criteria to attempt a brute force CW
structure inversion. Thus, we performed a structure inversion using the 1456
$J_{ij}$s within the 5 \AA~cut-off as independent free fitting parameters and
gradually increased the included number of first-principles calculations of the
energies of cells with random generated configurations $\boldsymbol{\sigma}$ in
the procedure. We found that when the number of considered configurations was
4000, the values of the MEI had converged to the MDCA obtained values, with a
mean absolute deviation of only 1.6 meV. The good agreement between the two
methods is illustrated in the inset in Fig.~\ref{fig:aCrN}. The MEI in amorphous
CrN are predominantly AFM, especially at short distances, but at larger
distances a fraction are ferromagnetic (FM).  For all pair
distances the spread in MEI is huge, underlying the need to treat all pair
interactions as independent in any quantitative modeling scheme of amorphous
magnets.

The obtained full set of exchange interactions up to $|\mathbf{r_i} -
\mathbf{r_j}| = 5$\AA ~were then used in the non-collinear Heisenberg Hamiltonian in Eq.~\ref{eq:heiss}.
We minimized the magnetic energy with respect to non-collinear configurations using a simulated annealing procedure with a Metropolis-type Monte Carlo simulation~\cite{Metropolis1953}. The energy was minimized in a 4x4x4 supercell of our amorphous structure model, including 8000 Cr magnetic moments coupled by the calculated exchange interactions. The system was initiated in a disordered non-collinear spin configuration followed by a gradual decrease in temperature with 100000 random trial spin rotations each temperature step. The procedure was initiated at several different starting temperatures in the interval 1000 - 100 K. The lowest energy state found was a non-collinear AFM configuration with a ferromagnetic component no larger than 0.01$\mu_B$ per Cr-atom. It was obtained for runs with starting temperatures in the range 200 - 100 K.



In conclusion we demonstrate the vital importance of local environment effects
on the magnetic exchange interactions in disordered systems, including amorphous
magnets. For this purpose we have suggested a supercell technique 
inspired by the direct cluster averaging method from alloy theory, which could
be fruitfully combined with a structure inversion approach. Our work opens up
for theoretical predictions of ground state ordering and finite temperature
magnetism in amorphous systems, as well as simulations of vibrational and
configurational local environment effects on magnetism in alloys and compounds. 



\begin{acknowledgments}
Financial support from the Swedish Research Council (VR) grants no.
621-2011-4417 and 621-2011-4426 and from the Swedish Foundation for Strategic Research (SSF) program SRLL10-0026 are gratefully acknowledged. EH and RL thank FONDECYT projects 1110602 and 1120334. Calculations were performed utilizing supercomputer resources supplied by the Swedish National Infrastructure for Computing (SNIC) at the PDC and NSC centers. 
\end{acknowledgments} 





\end{document}